# Image Encryption with Dynamic Chaotic Look-Up Table

Med Karim ABDMOULEH, Ali KHALFALLAH and Med Salim BOUHLEL
Research Unit: Sciences and Technologies of Image and Telecommunications
Higher Institute of Biotechnology
Sfax, Tunisia
medkarim.abdmouleh@isggb.rnu.tn; khalfallah.ali@laposte.net; medsalim.bouhlel@enis.rnu.tn

*Abstract*—**In this paper we propose a novel image encryption scheme. The proposed method is based on the chaos theory. Our cryptosystem uses the chaos theory to define a dynamic chaotic Look-Up Table (LUT) to compute the new value of the current pixel to cipher. Applying this process on each pixel of the plain image, we generate the encrypted image.**

**The results of different experimental tests, such as Key space analysis, Information Entropy and Histogram analysis, show that the proposed encryption image scheme seems to be protected against various attacks. A comparison between the plain and encrypted image, in terms of correlation coefficient, proves that the plain image is very different from the encrypted one.**

## I. Introduction

Nowadays, in the digital world, the security of transmitted digital images/videos becomes more and more vital, against the web attacks which become important [1]. So, cryptography is used to confirm security in open networks [2].

Cryptography is the science that uses mathematics to offer encryption algorithms to protect information.

Many cryptographic methods have been proposed. These techniques are classified into three categories:

- Symmetric cryptography: in this class the same key is used for encryption and decryption. Data Encryption Standard (DES) and Advanced Encryption Standard (AES) are the best known examples of symmetric encryption [3] [4].

- Asymmetric cryptography: two different keys are used. The first realization of a public key algorithm called RSA [5] according to its inventors Rivest, Shamir and Adelman, is the most used algorithm in asymmetric encryption.

- Hybrid cryptography: it is based on combination of the best features of symmetric cryptography and asymmetric [6].

Security has become a key issue in the world of electronic communication. An efficient cryptosystem should not be broken by hackers.

In this research we have tried to propose a secure and simple method of image encryption based on chaotic function.

This paper is organized as follows: section II describes the relation between chaos and cryptography and presents the Logistic Map function. The proposed method is introduced in section III. In section IV, we discuss the experimental results and test the efficiency of the algorithm. Finally, section V is devoted to the conclusion.

## II. Chaotic System

Since the 1990's, Chaotic Systems (CS) have attracted researchers to use them in secure communication [7] [8].

Most CS proposed in literature are characterized by certain properties [9] that facilitate their use in the design of modern cryptographic systems. These properties are essentially the ergodicity, sensitivity to the initial values and to the controlling parameters.

In modern cryptography, the CS have been used extensively in the development of cryptosystems. The CS are among the best known and most widely used in cryptography, they cite the Logistic Map (LM). The Logistic Map [10] is one of the most famous and simplest one-dimensional map. Therefore this function has been studied recently for cryptography applications. The LM is widely used in image encryption [11] [12].

The Logistic Map function is expressed by the following formula:

$$X_{n+1} = \mu\, X_n\, (1 - X_n) \qquad (1)$$

Where x is a floating number, which takes values in the interval [0, 1], n = 0, 1, 2, … and µ is the control parameter $0 < \mu \leq 4$.

If parameter µ takes a value between 0 and 3, the Logistic Map, which is defined by equation (1), is seen to converge to a particular value after some iteration. As parameter μ is further increased, the curves bifurcations become faster and faster. μ should be greater than 3.5784257 (known as the "point of accumulation") in order to have a chaotic system. Finally for μ=3.9 to 4, the chaos values are generated in the complete range of 0 to 1 [13].

## III. PROPOSED METHOD

In this section, we present our new image cryptosystem. The architecture of the proposed method is shown in "*Fig. 1-3*".

In the first step, we generate a chaotic matrix using the Logistic Map function $LM_1$ ($x_{0XOR}$ and $\mu_{0XOR}$). We mix this chaotic matrix with the plain image using the logical function XOR to obtain the initial encrypted image $I_1$.

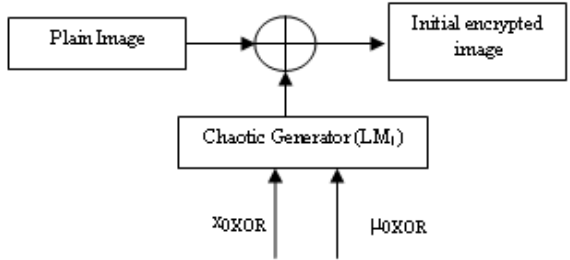

Figure 1. XOR Chaotic encryption

For each pixel $P_1$ from $I_1$, we generate a chaotic LUT using the Logistic Map function ($LM_2$) having as parameters $\mu_0$ and $x_0(P_c)$. Where $P_c$ is the value of the previous encrypted pixel by our cryptosystem. The initial condition $x_0(P_c)$ depends on the previous value of the ciphered pixel and $x_0$. In addition, we limit $x_0(P_c)$ values between 0.1 and 0.9.

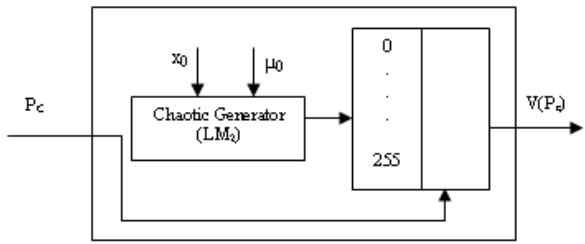

Figure 2. Chaotic Dynamic Look-Up Table

Applying the new LUT to $P_i$ we obtain the final encrypted pixel. We repeat this process for each pixel from the initial encrypted image to get the final encrypted one.

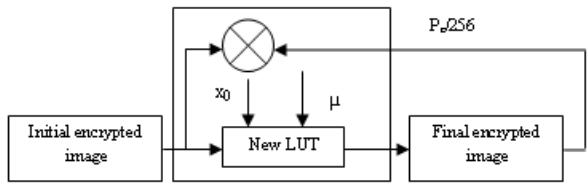

Figure 3. Chaotic Look-Up table encryption

## IV. EXPERIMENTAL RESULTS

Image information differs from the textual one and has very interesting properties such as its large capacity, redundancy and strong correlation strength between neighboring pixels. According to these properties, the security of a cryptosystem image is evaluated by a statistical analysis, the sensitivity analysis and the differential analysis [14] [15].

We test our method using an image test bank containing 30 different gray scale images with different size.

A good cryptosystem should ideally be robust against all kinds of attacks (cryptanalytic and statistical attacks). Thus, in this section we give the experimental results to prove the performance of our encryption image scheme.

### A. Key Space Analysis

The key used in our method includes the following combination {$x_0$, $\mu_0$, $x_{0XOR}$, $\mu_{0XOR}$} where all the parameters are real numbers. We use MATLAB 7.6 in our study. This mathematical tool codes real in 8 bytes. So, all the parameters are presented in 64 bits. Thus, the proposed encryption image scheme {$2^{64}$ x $2^{64}$ x $2^{64}$ x $2^{64}$} = $2^{256}$ has different combinations of the secret key.

### B. Histogram Analysis

Histogram analysis is one of the security analyses which give us the statistical properties of the ciphered image. Histogram may reflect how pixels in an image are distributed.

In order to have a perfect encrypted image, the histogram of the image must have a uniform distribution.

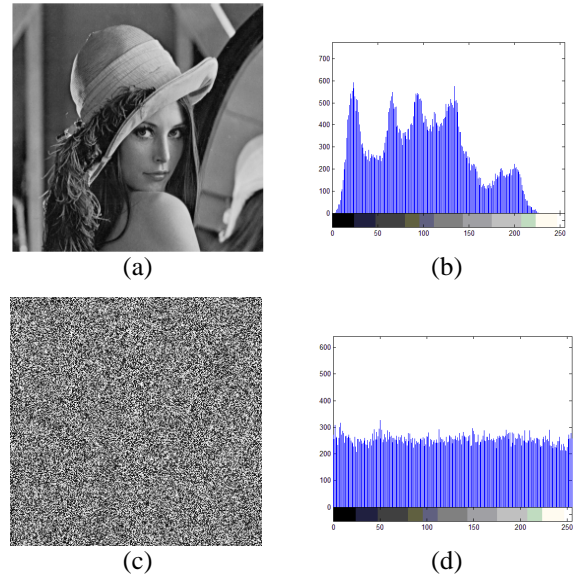

(a)    (b)

(c)    (d)

Figure 4. Histogram analysis: (a) plain image, (b) plain image histogram, (c) encrypted image, (d) encrypted image histogram

Referring to the obtained results, we can clearly see that the plain image "*Fig. 4.a*" differs significantly from the correspondent encrypted one "*Fig. 4.c*". Moreover, the histogram of the encrypted image "*Fig. 4.d*" is fairly uniform which makes it difficult to extract the pixels statistical nature of the plain image "*Fig. 4.b*".

## C. Information Entropy

The entropy was founded by Shannon in 1948 [16] [17] and is given in the following equation:

$$H(m) = \sum_i P(m_i) \log_2 \frac{1}{P(m_i)} \quad (2)$$

Where $P(m_i)$ represents the probability of symbol $m_i$. The entropy H(m) is expressed in bits.

From "*Fig. 4.d*", the encrypted image has a uniform histogram, which means that the gray levels have the same number of occurrences and hence the entropy is maximum. Therefore, a grayscale image, where each pixel is represented by 8 bits, must have an entropy for the encrypted image, the closest possible 8 bits/pixel. [18]

The entropy of the encrypted Lena image is equal to 7.9963 ≈ 8 bits/pixel. The obtained value is very close to the theoretical one. From this result, it is clear that our encryption image scheme is robust against the entropy attack.

## D. Correlation Coefficient Analysis

Another type of statistical analysis is the correlation coefficient analysis [19] [20] [21].

We calculate the correlation coefficient for a sequence of adjacent pixels using the following formula:

$$r_{xy} = \frac{\text{cov}(x, y)}{\sqrt{D(x)}\sqrt{D(y)}} \quad (3)$$

Here, x and y are the intensity values of two adjacent pixels in the image. $r_{xy}$ is the correlation coefficient. The cov(x,y), E(x) and D(x) are given as follows:

$$E(x) = \frac{1}{N} \sum_{i=1}^{N} x_i \quad (4)$$

$$D(x) = \frac{1}{N} \sum_{i=1}^{N} [x_i - E(x_i)]^2 \quad (5)$$

$$\text{cov}(x, y) = \frac{1}{N} \sum_{i=1}^{N} [(x_i - E(x_i))(y_i - E(y_i))] \quad (6)$$

N is the number of adjacent pixels selected from the image to calculate the correlation.

The table below presents the correlation between two adjacent pixels for the plain and the encrypted Lena image.

To test the correlation coefficient, we have chosen 2500 pairs of two adjacent pixels which are selected randomly from both the plain and encrypted image.

TABLE I. CORRELATION COEFFICIENTS OF TWO ADJACENT PIXELS IN THE PROPOSED METHOD

| Direction | Correlation plain image | Correlation encrypted image |
|---|---|---|
| Diagonal | 0.9177 | 0.0027 |
| Horizontal | 0.9399 | -0.0214 |
| Vertical | 0.9692 | -0.0022 |

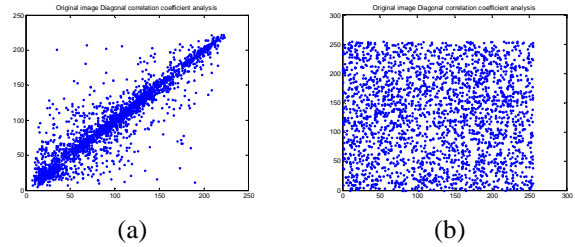

(a)  (b)

Figure 5. Correlation between two diagonally adjacent pixels: (a) in the plain image, (b) in the encrypted image

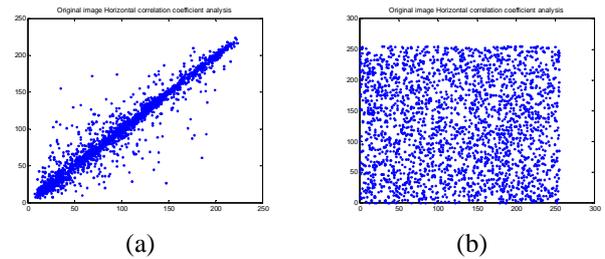

(a)  (b)

Figure 6. Correlation between two horizontally adjacent pixels: (a) in the plain image, (b) in the encrypted image

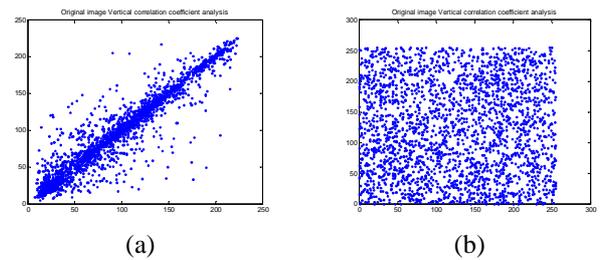

(a)  (b)

Figure 7. Correlation between two vertically adjacent pixels: (a) in the plain image, (b) in the encrypted image

*Fig. 5-7* represents the correlation between two diagonally, horizontally and vertically adjacent pixels of the plain and encrypted image. We see that the neighboring pixels in the plain image have a high correlation, while in the encrypted one there is a little correlation. This little correlation between two neighbors pixels in the ciphered image makes the brook of our cryptosystem difficult.

## E. Differential Analysis

Three criteria NPCR, UACI and MAE [22] are used to test the sensitivity of a single-bit change in the plain image.

The NPCR means the Number of Pixels Change Rate of ciphered image while a pixel of the plain image is changed.

Let $A_2$ be the changed plain image on one pixel. $C_1$ and $C_2$ are the ciphered images of the plain image and $A_2$. D is a matrix having the same size as the image figures $C_1$ and $C_2$. $D(i, j)$ is determined as follows:

$$D(i, j) = \begin{cases} 1 & \text{if } C_1(i, j) \neq C_2(i, j) \\ 0 & \text{else} \end{cases} \quad (7)$$

The NPCR is defined by:

$$NPCR = \frac{\sum_{i=0}^{M-1} \sum_{j=0}^{N-1} D(i, j)}{M \times N} \times 100 \quad (8)$$

M and N are the height and width of encrypted images $C_1$ and $C_2$.

The Unified Average Changing Intensity (UACI) between these two images is defined by the following formula:

$$UACI = \frac{1}{M \times N} \sum_{i=0}^{M-1} \sum_{j=0}^{N-1} \frac{|C_1(i, j) - C_2(i, j)|}{255} \times 100 \quad (9)$$

The Mean Absolute Error is defined by the following equation:

$$MAE = \frac{1}{M \times N} \sum_{i=0}^{M-1} \sum_{j=0}^{N-1} |C_1(i, j) - C_2(i, j)| \quad (10)$$

The NPCR, UACI and MAE results are given in Table II.

TABLE II. VALUES RESULTS OF NPCR, UACI AND MAE

|  | NPCR | UACI | MAE |
|---|---|---|---|
| Lena | 99.2874 | 31.5689 | 80.5007 |

Experimentally measured value of NPCR is 99.2874 % and UACI is 31.5689 % for Lena image. These results indicate that a small change in the plain image introduces a high alteration on the ciphered one. So, the proposed method resists the differential attacks.

## F. Key Sensitivity Test

The secret key of our encryption algorithm is composed of four values {$x_0$, $\mu_0$, $x_{0XOR}$, $\mu_{0XOR}$}. These values are the parameters of the two Logistic Map functions used in our cryptosystem. Thus, to analyze the sensitivity of our approach to a little change in the secret key, we encrypt Lena image with key $k_0$, then, we encrypt the same image with keys $k_1$, $k_2$, $k_3$ and $k_4$ which present slight differences from the key $k_0$ as shown in table III.

TABLE III. THE KEYS USED IN THE EXPERIMENTS

|  | $k_0$ | $k_1$ | $k_2$ | $k_3$ | $k_4$ |
|---|---|---|---|---|---|
| $x_0$ | 0.4 | 0.4+10$^{-15}$ | 0.4 | 0.4 | 0.4 |
| $\mu_0$ | 3.9 | 3.9 | 3.9+10$^{-15}$ | 3.9 | 3.9 |
| $x_{0XOR}$ | 0.5002 | 0.5002 | 0.5002 | 0.5002+10$^{-15}$ | 0.5002 |
| $\mu_{0XOR}$ | 3.87001 | 3.87001 | 3.87001 | 3.87001 | 3,87001+10$^{-15}$ |

We label $I_C$ ("*Fig. 8.b*") the encrypted image of the plain image I ("*Fig. 8.a*") by our encryption image scheme using the key $k_0$. $I_{C1}$, $I_{C2}$, $I_{C3}$ and $I_{C4}$ are the encrypted images of the plain image I using respectively the keys $k_1$, $k_2$ ("*Fig. 9.a-b*"), $k_3$ and $k_4$ ("*Fig. 10.a-b*").

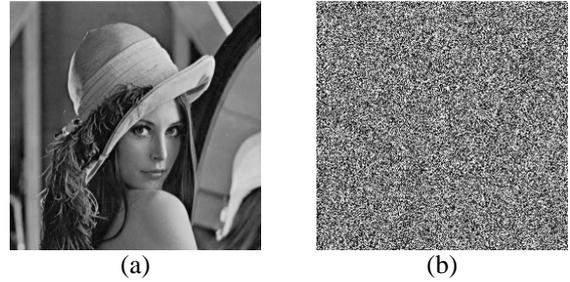

(a)      (b)

Figure 8. (a) Plain image Lena, (b) encrypted image utilizing the original key $k_0$

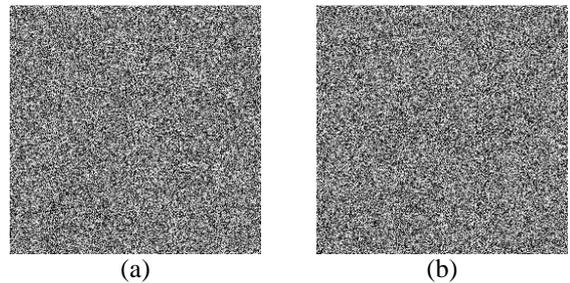

(a)      (b)

Figure 9. (a) Encrypted image utilizing the key $k_1$, (b) encrypted image utilizing the key $k_2$

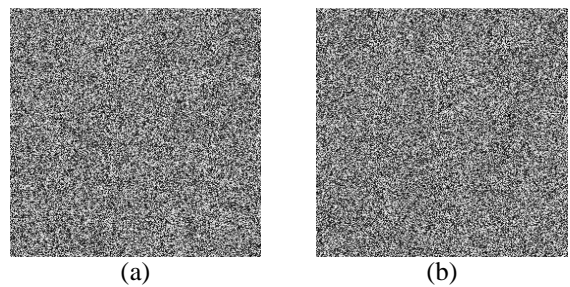

(a)      (b)

Figure 10. (a) Encrypted image utilizing the key $k_3$, (b) encrypted image utilizing the key $k_4$

Visually, we can not prove any difference between the ciphered images encrypted with different keys. In fact, the use of the Logistic Map in the encryption schemes introduces a great alteration on the plain image. But according to tables belows (Table IV, Table V and Table VI) which summarize the comparison between the different images on terms of NPCR, MAE and UACI, prove that these images are very different. In fact, all NPCR values are higher than 99 % and the UACI values are greater than 30 %. We can deduce that our encryption algorithm is sensitive to the secret key.

TABLE IV.    *NPCR TABLE FOR SENSITIVITY TO CHANGING CIPHERING KEY*

|  | $I_C$ | $I_{C1}$ | $I_{C2}$ | $I_{C3}$ | $I_{C4}$ | I |
|---|---|---|---|---|---|---|
| $I_C$ | 0 | 99.26 | 99.33 | 99.63 | 99.59 | 99.59 |
| $I_{C1}$ | 99.26 | 0 | 99.25 | 99.26 | 99.26 | 99.60 |
| $I_{C2}$ | 99.33 | 99.25 | 0 | 99.33 | 99.33 | 99.59 |
| $I_{C3}$ | 99.63 | 99.26 | 99.33 | 0 | 99.60 | 99.60 |
| $I_{C4}$ | 99.59 | 99.26 | 99.33 | 99.60 | 0 | 99.63 |
| I | 99.59 | 99.60 | 99.59 | 99.60 | 99.63 | 0 |

TABLE V.    *MAE TABLE FOR SENSITIVITY TO CHANGING CIPHERING KEY*

|  | $I_C$ | $I_{C1}$ | $I_{C2}$ | $I_{C3}$ | $I_{C4}$ | I |
|---|---|---|---|---|---|---|
| $I_C$ | 0 | 80.47 | 80.52 | 85.04 | 84.93 | 76.55 |
| $I_{C1}$ | 80.47 | 0 | 80.47 | 80.47 | 80.47 | 76.51 |
| $I_{C2}$ | 80.52 | 80.47 | 0 | 80.52 | 80.52 | 76.69 |
| $I_{C3}$ | 85.04 | 80.47 | 80.52 | 0 | 85.32 | 76.78 |
| $I_{C4}$ | 84.93 | 82.71 | 80.52 | 85.32 | 0 | 76.36 |
| I | 76.55 | 76.51 | 76.69 | 76.78 | 76.36 | 0 |

TABLE VI.    *UACI TABLE FOR SENSITIVITY TO CHANGING CIPHERING KEY*

|  | $I_C$ | $I_{C1}$ | $I_{C2}$ | $I_{C3}$ | $I_{C4}$ | I |
|---|---|---|---|---|---|---|
| $I_C$ | 0 | 31.56 | 31.58 | 33.35 | 33.30 | 30.02 |
| $I_{C1}$ | 31.56 | 0 | 31.56 | 31.56 | 32.44 | 30.00 |
| $I_{C2}$ | 31.58 | 31.56 | 0 | 31.58 | 31.58 | 30.07 |
| $I_{C3}$ | 33.35 | 31.56 | 31.58 | 0 | 33.46 | 30.11 |
| $I_{C4}$ | 33.30 | 31.56 | 31.58 | 33.46 | 0 | 29.95 |
| I | 30.02 | 30.00 | 30.07 | 30.11 | 29.95 | 0 |

Thus, based on the obtained results, we can conclude that a slight change in the encryption key (order of $10^{-15}$ for real {$x_0$, $\mu_0$, $x_{0XOR}$, $\mu_{0XOR}$}) we get two totally different encrypted images.

We now propose to analyze the secret key sensitivity of our approach on the decryption scheme. To do this, we encrypted a clear image with key $k_0$, and we decrypted it with other keys, which are slightly different from the encryption key. $I_D$, $I_{D1}$, $I_{D2}$, $I_{D3}$ and $I_{D4}$ are the deciphered of $I_C$ using the keys $k_0$, $k_1$, $k_2$, $k_3$ and $k_4$. Visually, we remark that, using a false deciphering key yields to a decipher failure "*Fig. 11.a-e*".

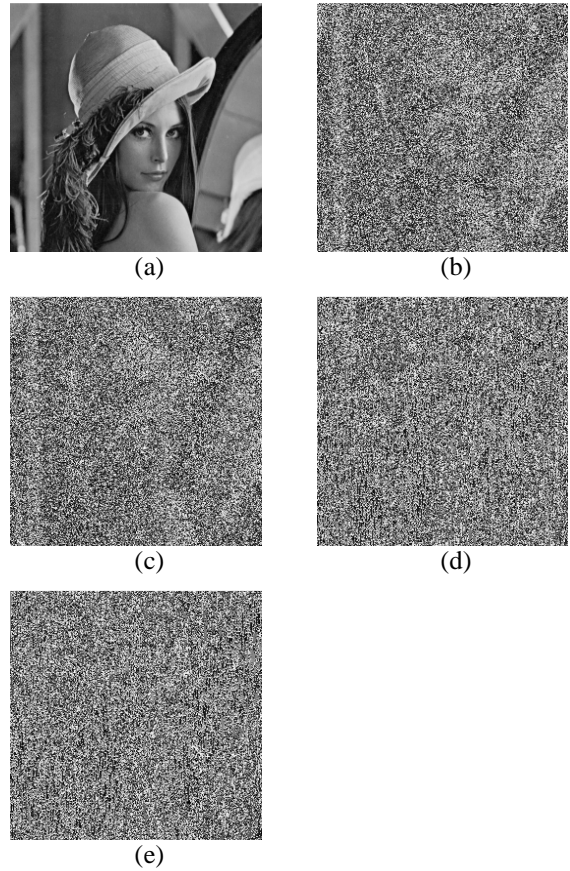

Figure 11. Key sensitivity test: (a) $I_D=I_C$ deciphered image using $k_0$, (b) $I_{D1}=I_C$ deciphered image using $k_1$, (c) $I_{D2}=I_C$ deciphered image using $k_2$, (d) $I_{D3}=I_C$ deciphered image using $k_3$, e) $I_{D4}=I_C$ deciphered image using $k_4$

The tables below (Table VII, Table VIII and Table IX) indicate respectively the values of NPCR, MAE and UACI obtained for images with very close decrypted keys.

TABLE VII.    *NPCR TABLE FOR SENSITIVITY TO CHANGING DECIPHERING KEY*

|  | $I_D$ | $I_{D1}$ | $I_{D2}$ | $I_{D3}$ | $I_{D4}$ | I |
|---|---|---|---|---|---|---|
| $I_D$ | 0 | 99.28 | 99.23 | 98.77 | 98.83 | 0 |
| $I_{D1}$ | 99.28 | 0 | 99.20 | 99.60 | 99.58 | 99.28 |
| $I_{D2}$ | 99.23 | 99.20 | 0 | 99.60 | 99.59 | 99.23 |
| $I_{D3}$ | 98.77 | 99.60 | 99.60 | 0 | 98.86 | 98.77 |
| $I_{D4}$ | 98.83 | 99.58 | 99.59 | 98.86 | 0 | 98.83 |
| I | 0 | 99.28 | 99.23 | 98.77 | 98.83 | 0 |

TABLE VIII. MAE TABLE FOR SENSITIVITY TO CHANGING DECIPHERING KEY

|  | $I_D$ | $I_{D1}$ | $I_{D2}$ | $I_{D3}$ | $I_{D4}$ | I |
|---|---|---|---|---|---|---|
| $I_D$ | 0 | 67.72 | 67.92 | 75.41 | 75.30 | 0 |
| $I_{D1}$ | 67.72 | 0 | 73.47 | 83.92 | 84.25 | 67.72 |
| $I_{D2}$ | 67.92 | 73.47 | 0 | 83.87 | 84.26 | 67.92 |
| $I_{D3}$ | 75.41 | 83.92 | 83.87 | 0 | 82.01 | 75.41 |
| $I_{D4}$ | 75.30 | 84.25 | 84.26 | 82.01 | 0 | 75.3 |
| I | 0 | 67.72 | 67.92 | 75.41 | 75.3 | 0 |

TABLE IX. UACI TABLE FOR SENSITIVITY TO CHANGING DECIPHERING KEY

|  | $I_D$ | $I_{D1}$ | $I_{D2}$ | $I_{D3}$ | $I_{D4}$ | I |
|---|---|---|---|---|---|---|
| $I_D$ | 0 | 26.56 | 26.63 | 29.57 | 29.53 | 0 |
| $I_{D1}$ | 26.56 | 0 | 28.81 | 32.91 | 33.04 | 26.56 |
| $I_{D2}$ | 26.63 | 28.81 | 0 | 32.89 | 33.04 | 26.63 |
| $I_{D3}$ | 29.57 | 32.91 | 32.89 | 0 | 32.16 | 29.57 |
| $I_{D4}$ | 29.53 | 33.04 | 33.04 | 32.16 | 0 | 29.53 |
| I | 0 | 26.56 | 26.63 | 29.57 | 29.53 | 0 |

The obtained results in tables VII-IX boost the visual inspection and prove that a slight error introduced on the true encryption key yields to the image decryption failure. Consequently, our new encoding image scheme is highly sensitive to any change in the ciphering key.

G. *Cryptanalysis*

While cryptography is the discipline that ensures the security of confidential information, cryptanalysis [3] is the discipline that studies and validates the robustness of cryptosystems against attacks. According to Kerckhoffs, the security of a cryptosystem depends on the secrecy of the key and not on that of the encryption algorithm. To study the cryptosystem security, we can utilize the Kerckhoffs principle. In this case, the cryptanalyst must be unable to find the key even if he has access to the plaintext and its corresponding ciphertext. He tries to apply more attacks such as:

- Ciphertext only attack
- Chosen plaintext attack
- Known plaintext attack
- Chosen ciphertext attack

Known-plaintext attack is one of the attacks where the cryptanalyst owns a plaintext image and its corresponding encrypted image. This frequently used attack utilizes the known clear image and the ciphered image to extract the decryption key or to decrypt another ciphered image. In our study, we are based on the known plaintext attack to apply the stream key attack on our cryptosystem. "*Fig. 12*" summarizes the key stream attack applied on our cryptosystem. *Fig. 12.c* represents the key stream used to decrypt the ciphered image "*Fig. 12.b*". The result of this process is shown in "*Fig. 12.d*". We remarks that the obtained image is equal to the plain image "*Fig. 12.a*". After that, we utilize the extracted key stream to decrypt the image shown in "*Fig. 12.e*", which represents the encryption of the "Goldhill" image. *Fig. 12.f* proves the failure of this attack. In fact, this last image is fairly different from the "Goldhill" plain image illustrated in "*Fig. 12.g*".

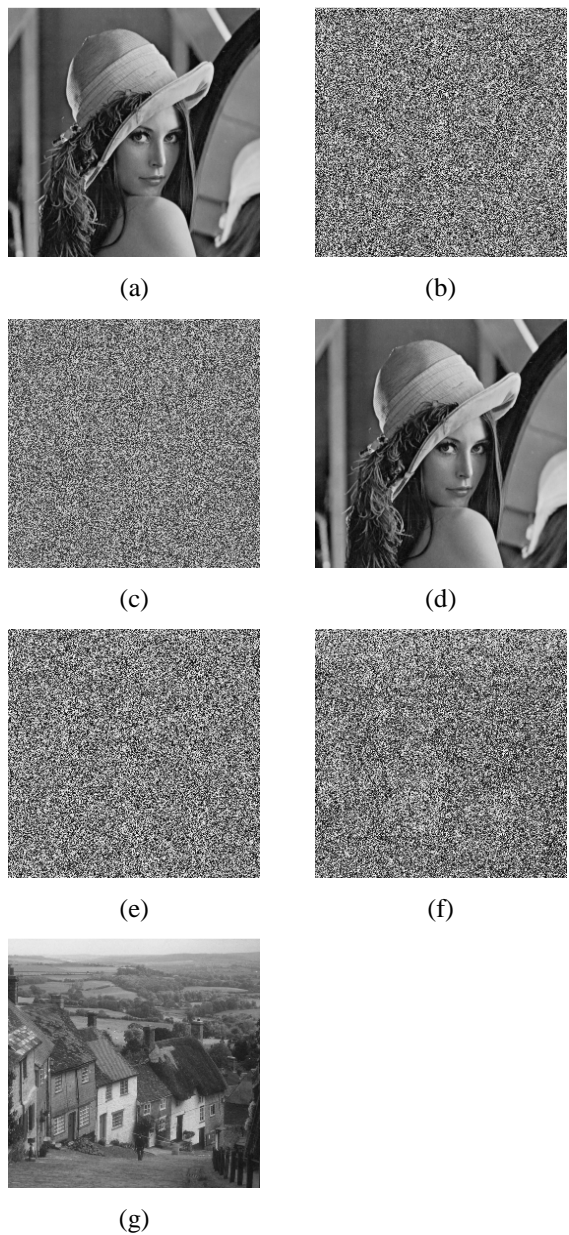

Figure 12. Failed crack attempt:
(a) plain image Lena, (b) encrypted Lena image, (c) extracted key stream, (d) decrypted Lena image, (e) encrypted Goldhill image, (f) failed attempt to crack the cipher image of Goldhill, (g) plain image Goldhill

The established results shown in "*Fig. 12*" demonstrate that our approach could not be broken by a known plaintext text attack.

## V. Conclusion

In this paper, we have proposed an image encryption scheme based on the chaos theory. This new proposal uses the Logistic Map function to generate a dynamic LUT. This LUT depends on previous encrypted pixel. This Feedback introduced on the cryptosystem gives it excellent performance. In fact, the encrypted image is very different from the plaintext one. This dissimilarity is due to the random introduced by the chaos. The use of the Logistic Map gives the proposed cryptosystem a sensitivity to the encryption and decryption keys. Finally we have tested our cryptosystem against the known plaintext attack. The obtained results prove that our cryptosystem could not be broken.